\documentclass[aps,prx,english,
,reprint]{revtex4-1}
\usepackage[utf8]{inputenc}
\usepackage{longtable}
\usepackage{morefloats}
\usepackage{tabularx}
\usepackage{multirow}
\usepackage[dvips]{graphicx}
\usepackage{color}
\usepackage{epsfig,amsbsy}
\usepackage{amsmath,amsfonts,amsthm,amssymb}
\usepackage{appendix}
\usepackage{textcomp}
\usepackage{makeidx}
\usepackage{epstopdf}
\usepackage{url}
\usepackage[bookmarksnumbered,pdfpagelabels=true,plainpages=false,colorlinks=true,linkcolor=blue,citecolor=red,urlcolor=blue]{hyperref}
\usepackage{ulem}
\usepackage{xparse}
\usepackage{tikz}
\usetikzlibrary{matrix,backgrounds}
\pgfdeclarelayer{myback}
\pgfsetlayers{myback,background,main}
\usepackage{float}
\usepackage{mathtools} 
\let\mbf\mathbf
\let\t\text

\tikzset{mycolor/.style = {line width=0.1bp,color=#1}}%
\tikzset{myfillcolor/.style = {draw,fill=#1}}%

\NewDocumentCommand{\fhighlight}{O{blue!40} m m}
{\draw[rounded corners,ultra thin, fill=red,draw=black, opacity=0.1] (m-1-1.north west)rectangle (m-2-2.south east);
\draw[rounded corners,ultra thin, fill=red,draw=black, opacity=0.1] (m-3-3.north west)rectangle (m-4-4.south east);
\draw[rounded corners,ultra thin, fill=blue,draw=black, opacity=0.03] (m-1-3.north west)rectangle (m-2-4.south east);
\draw[rounded corners,ultra thin, fill=blue,draw=black, opacity=0.03] (m-3-1.north west)rectangle (m-4-2.south east);}

\begin{document}

\title{
Dzyaloshinskii-Moriya Interaction and Dipole-Exchange Curvature Effects on the Spin-Wave Spectra of Magnetic Nanotubes}

\author{B. Mimica-Figari}
\affiliation{Departamento de F\'{i}sica, Universidad T\'{e}cnica Federico Santa Mar\'{i}a, Avenida Espa\~{n}a 1680, Valpara\'{i}so, Chile}
\author{P. Landeros}
\affiliation{Departamento de F\'{i}sica, Universidad T\'{e}cnica Federico Santa Mar\'{i}a, Avenida Espa\~{n}a 1680, Valpara\'{i}so, Chile}
\author{R. A. Gallardo}
\email{rodolfo.gallardo@usm.cl}
\affiliation{Departamento de F\'{i}sica, Universidad T\'{e}cnica Federico Santa Mar\'{i}a, Avenida Espa\~{n}a 1680, Valpara\'{i}so, Chile}
\date{\today }

\begin{abstract}

This work explores spin-wave dynamics in magnetic nanotubes, focusing on the influence of the Dzyaloshinskii-Moriya interaction and curvature. The study uses analytical methods to examine how these factors influence the emergence of nonreciprocity and azimuthal standing waves in nanotubes with longitudinal magnetization along the axis or with a vortex-like magnetization. The interplay between exchange, Dzyaloshinskii-Moriya, and dipolar couplings in determining the chirality of spin waves is discussed. When the magnetization is saturated along an axis, the spin waves propagating along it are symmetric under the inversion of the wave vector. However, magnetochirality, mainly driven by exchange and Dzyaloshinskii-Moriya couplings, is observed in the azimuthal standing modes. In the vortex state, frequency nonreciprocity occurs for waves propagating along the tube, while the azimuthal modes remain reciprocal. For positive Dzyaloshinskii-Moriya interaction, and depending on the helicity of the vortex, the asymmetry induced by the dipolar interaction is reinforced, whereas a negative coupling opposes this asymmetry. The influence of radial anisotropy is also examined. It is found that radial anisotropy reduces the frequency of the modes and shifts the dispersion minimum to a finite wave vector in the vortex state. The properties of modes near zero frequency offer insight into the emergence of chiral magnetic textures. 

\end{abstract}

\maketitle

\section{Introduction}
\label{sec:introduction}

Spin-wave (SW) nonreciprocity is a phenomenon in which the propagation characteristics of two counterpropagating waves differ \cite{Barman21,Flebus24}. These waves may exhibit amplitude, phase, attenuation length, or frequency variations, even when their wave vectors have the same magnitude \cite{Jamali13,Albisetti20,Pirro21}. 
In thick films, an asymmetry in the SW amplitude occurs for the Damon-Eshbach modes \cite{Damon61}, where waves propagating in opposite directions are localized on the top or bottom surface of the film. This effect is given mainly in the small wave vector range, where the dipolar interaction dominates. Such nonreciprocity in amplitude results from the asymmetric distribution of the internal dipolar fields throughout the thickness of the film, which causes the magnetization dynamics to vary between the top and bottom surfaces \cite{Gladii16,Solano22,Szulc24}.
The dipolar asymmetry is relevant for films thicker than a few exchange lengths, which is the subject of three-dimensional nanomagnetism \cite{Gubbiotti19,Gubbiotti25}.

Nonreciprocity in frequency driven by the dipole-dipole interaction has been observed in various nanostructures, including magnetization-graded films \cite{Gallardo19a}, synthetic antiferromagnets \cite{Grunberg85,Dib15,Gallardo19b,Franco20,Grassi20,Gladii23,Gallardo24,Girardi24}, and magnetic nanotubes \cite{Otalora16,Salazar21,Sheka20,Gallardo22a,Korber22}. It has also been observed in films with asymmetric surface anisotropies \cite{Gladii16,Solano22,Szulc24} and antisymmetric exchange known as the Dzyaloshinskii-Moriya interaction (DMI) \cite{Dzyaloshinskii58,Moriya60,Fert80}. The DMI influence on the SW spectra has been theoretically predicted \cite{Melcher73,Costa10,Cortes13,Moon13} and experimentally observed in ultrathin films \cite{Kai15,Cho15,Stashkevich15,Tacchi17b,Gallardo19BCh,Kuepferling23} and noncentrosymmetric materials \cite{Iguchi15,Seki16}. Unlike the parallel spin alignment favored by ferromagnetic exchange interactions, DMI promotes a noncollinear alignment of neighboring spins. This property induces the asymmetry in the SW dispersion but also leads to the formation of chiral spin textures \cite{Rohart13,Franken14}. Consequently, the DMI plays a crucial role in engineering and controlling spin waves in magnonic logic devices, paving the way for applications in nanoscale information processing \cite{Lee16,Chumak17,LiYi20,Wang24}.

Three-dimensional nanomagnetism \cite{Gubbiotti19,Gubbiotti25} deals with complex systems and planar and curvilinear shells \cite{Sheka21,Makarov22a} with a thickness larger than some tens of nanometers. In the case of ferromagnetic nanotubes (NTs) \cite{Stano18,Landeros22}, the interplay between dipolar and exchange couplings can lead to distinctive magnetic textures \cite{Landeros09,Ruffer12,Mehlin18, Vasyukov18,Yang18,Landeros22} and domain walls \cite{Landeros10,Yan12,Otalora13,Hertel16}. 
For tubes hosting a vortex magnetization, spin-wave nonreciprocity is influenced by the curvature, where geometric charges induce a chiral dipolar interaction \cite{Sheka20}. This direction-dependent dipole-dipole coupling can lead to different frequencies \cite{Otalora16} or attenuation lengths for spin waves traveling in opposite directions \cite{Otalora18}. 
Like in the synthetic antiferromagnets \cite{Gallardo19b}, the SW asymmetry induced by dipolar interaction in nanotubes depends on the direction of the equilibrium magnetization and the SW propagation. Namely, nonreciprocity is observed in cases where the waves propagate perpendicular to the equilibrium magnetization, as can be determined by calculating the texture-dependent toroidal moment \cite{Korber22,Brevis24a}. In the tubular case, this is optimally reached when the magnetization describes a vortex state while the SWs propagate along the axis \cite{Salazar21}.  
Since the cooperative dipolar coupling is amplified in thicker shells, the spin-wave asymmetry enhances as the nanotube thickness increases \cite{Gallardo22a, Korber22}. Because curvature is the primary source of the nonreciprocal dipolar interaction, a reduction in curvature leads to a corresponding decrease in spin-wave asymmetry \cite{Brevis24b}. An approach to achieving SW asymmetries from interlayer dynamic coupling and curvature involves connecting two nanotubes in an antiparallel configuration, referred to as cylindrical synthetic antiferromagnets \cite{Gallardo22c}. In this arrangement, which combines the features of curvilinear shells and bilayers, nonreciprocity persists even if the curvature of the nanotubes diminishes.

During the curling transition from a fluxless vortex to an axially saturated magnetic state \cite{Gutierrez17}, it has been shown that the exchange interaction also induces chiral effects in the SW bandstructure \cite{Salazar21}. Here, the application of magnetic fields allows for tuning the magnon nonreciprocity, which can shift between dipolar- and exchange-dominated regimes based on the equilibrium state of magnetization \cite{Salazar21}. In the curling state \footnote{We prefer to call it curling instead of helical state, to avoid confusion with the DMI-induced helical states.}, where the magnetization is oriented at an oblique angle $\theta$ to the nanotube axis, the interplay between dipolar and exchange interactions generates distinct dispersions for various combinations of azimuthal and axial wave vectors. 
Experimental evidence of these predictions has been obtained in a nanotube with a hexagonal cross-section \cite{Giordano23} and in artificial chiral magnets with left or right-handed curling magnetizations \cite{MXu24}.
In the axial magnetic state, nodal lines were found to be mainly straight along the circumference. In contrast, for curling and circumferential magnetization orientations, the nodal lines exhibited distortions, which were attributed to the magnetochiral effects. This behavior reveals an unconventional nature of confined modes, akin to those observed in planar magnets under the influence of the DMI \cite{Zingsem19,Gallardo19c,Flores22,Gallardo24b}. 

Beyond the symmetric exchange, understanding the role of additional spin coupling in tubular systems is of high interest, particularly the interfacial DMI, which can be introduced by coating a nanotube with a high spin-orbit coupling heavy-metal (HM) layer \cite{Kuepferling23,Gallardo19BCh}. This interaction promotes the formation of chiral magnetic textures \cite{Avci19,Emori13} and drives spin-wave nonreciprocity with significant implications for spintronic applications \cite{Barman21,Flebus24}. Despite its fundamental importance, the influence of DMI on spin-wave dynamics in tubular nanostructures remains unexplored from an analytical point of view. This gap is likely due to the challenges associated with incorporating dynamic chiral terms into the unique tubular geometry. Only numerical addressing has been realized, where it is observed that both the curvature-induced magnetic charges and the bulk DMI may contribute to a dispersion asymmetry \cite{Korber21c}. 

This paper analytically investigates spin-wave dynamics in magnetic nanotubes with an interfacial Dzyaloshinskii-Moriya interaction. It examines how the DMI influences nonreciprocal spin-wave modes and azimuthal standing waves for different magnetization configurations (longitudinal and vortex-like). Using a theoretical model that incorporates Zeeman, exchange, dipolar interactions, radial anisotropy, and DMI, the study emanates a dispersion relation that, at a small wave vector limit, allows deriving an analytical expression for the frequency shift that quantifies the spin-wave asymmetry. Additionally, the influence of radial anisotropy is explored, where insights into the formation of chiral magnetic textures under specific conditions are discussed.

\section{Theoretical model}
\label{Theoretical model}

\begin{figure}[t]
  \includegraphics[width=.4\textwidth]{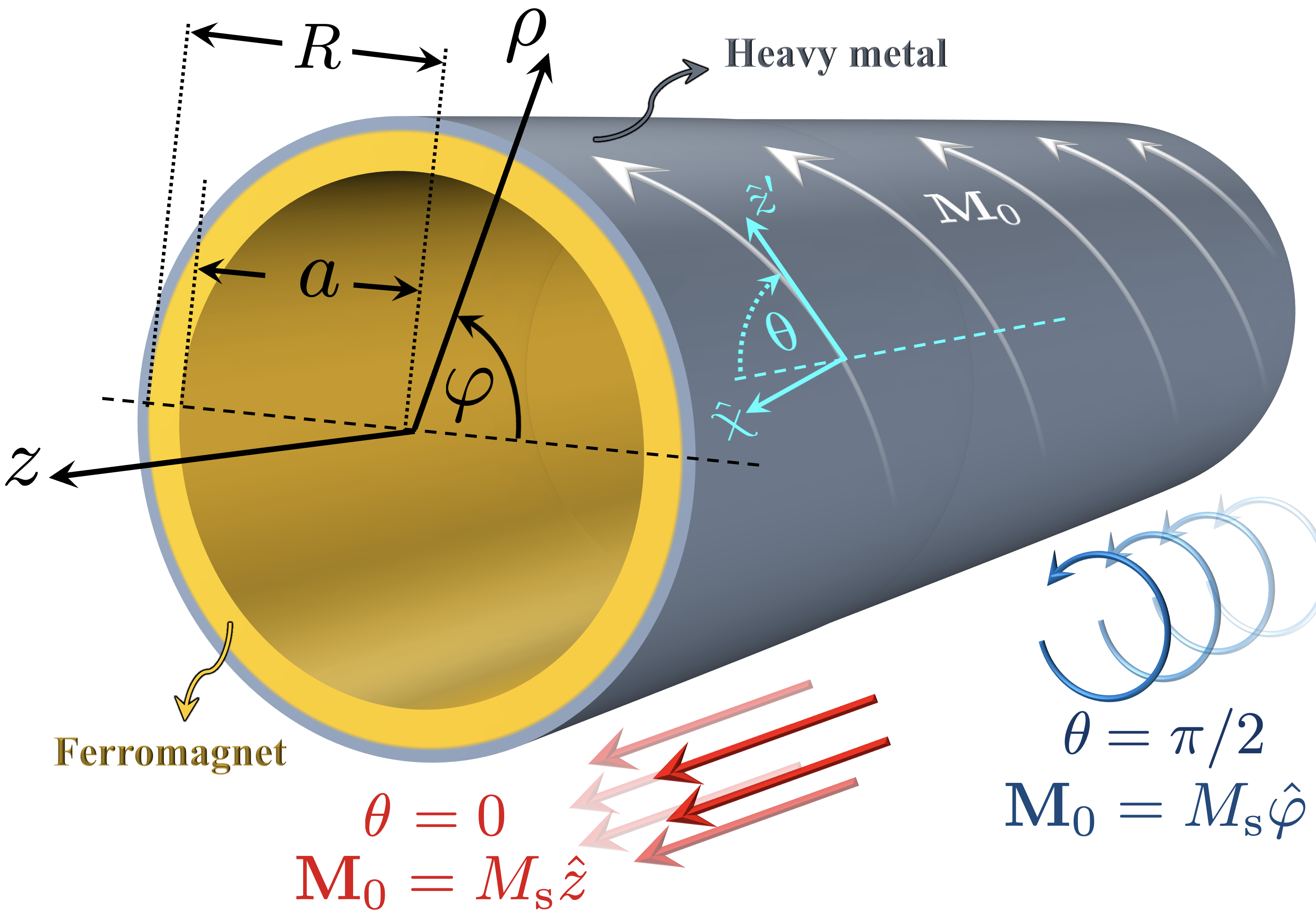}
  \caption{Ferromagnetic nanotube with an equilibrium magnetization $\mathbf{M}_0$ describing a curling state. The cylindrical ferromagnet of outer (inner) radius $R$ ($a$) is covered by a heavy-metal layer provoking an interfacial DMI. A glogal reference coordinate system ($\rho$,$\varphi$,$z$) is defined, where $z$ is oriented along the tube axis. A local reference is also defined ($\rho$,$\chi$,$z'$), where the equilibrium magnetization $\mbf M_0$ points along $\hat{z}'$, which makes an angle $\theta$ with the $z$ axis. The cases $\theta=0$ and $\theta=\pi/2$ are illustrated below, which describe an axial and vortex-like configuration, respectively.}
  \label{FIG1}
\end{figure}

A theoretical model is developed to describe the spin-wave dynamics in a nanotube with curling equilibrium magnetization, considering the contributions of Zeeman, exchange interaction, dipolar interaction, radial anisotropy, and the interfacial DMI. While the complete transition from a vortex to a longitudinal magnetized configuration can be theoretically explored, the results of this work will only focus on the limiting cases $\theta=0,\pi/2$. Nevertheless, the expressions for the static and dynamic effective fields and the equilibrium condition are written in a general way for completeness. 
The temporal evolution of the magnetization is given by the Landau-Lifshitz (LL) equation \cite{Gurevich96},
\begin{equation}
    \frac{d\mbf M(\mbf r,t)}{dt}  = -\mu_0 \gamma \mbf M(\mbf r,t)\times\mbf H^{\rm e}(\mbf r,t).
    \label{LL}
\end{equation}
where $\mbf M (\mbf r,t)$ is the magnetization, $\mbf H^{\rm e}(\mbf r,t)$ the effective magnetic field, and $\gamma$ the gyromagnetic ratio. Considering the wave propagation along the tube axis $z$ and assuming small magnetization deviation, the magnetization and the effective field are, respectively, given by $\mbf M(\mbf r,t) = \mbf M_0 +\mbf m e^{i(k z+l\varphi-\omega t)}$ and $\mbf H^{\rm e}(\mbf r,t) = \mbf H^{\rm e0} +\mbf h^{\rm e} e^{i(k z+l\varphi-\omega t)}$, where $\vert \mbf M_0\vert \gg \vert \mbf m\vert$ and $\vert \mbf H^{\rm e0}\vert \gg \vert \mbf h^{\rm e}\vert$, with $\mbf M_0$ being the equilibrium magnetization and $\mbf H^{\rm e0}$ the static effective field. 
Here, $\omega=2\pi f$ is the angular frequency, $l$ corresponds to an integer number that describes the azimuthal standing wave modes, and $k$ is the axial component of the wave vector. 
The equilibrium magnetization is considered to be in the direction $\hat{ z}' =  \mathcal C\sin\theta\hat{\mbf \varphi}+\cos\theta \hat{ z}$. Thus, for $\theta = 0$, the magnetization is saturated along the tube axis $z$, and for $\theta =\pi/2$, the magnetization presents a curling or vortex-like equilibrium state, as shown in Fig.~\ref{FIG1}. The chirality parameter \(\mathcal C \) has been added to represent the two possible curling orientations, namely \(\mathcal C = 1\) for a right-handed and \(\mathcal C  = -1\) for a left-handed vortex. The effective field contains all the interactions of the system. 
The magnetic oscillations are described through a local reference system defined by the plane orthogonal to the equilibrium magnetization. Such a coordinate system is constructed by considering that the equilibrium magnetization is orthogonal to the radial direction, i.e., $\hat{\mbf \rho}\cdot\hat{ z}' = 0$. Then, the right-handed orthonormal set is given by $ (\hat{\mbf\chi}, \hat{\mbf\rho},\hat{ z}')$, where $\hat\chi =  \hat{\mbf\rho} \times \hat{ z}'  = -\cos\theta \hat{\mbf \phi} + \mathcal C \sin\theta \hat{\mbf z}$. In terms of the magnetization components, the first-order Landau-Lifshitz equation becomes 
    \begin{subequations}
\label{LL des}
\begin{gather}
    \omega m_\chi = i \gamma \mu_0 (-H^{\rm e0}_{z'} m_\rho + M_{\t s} h^{\rm e}_\rho)\\
    \omega m_\rho = i \gamma\mu_0(H^{\rm e0}_{z'} m_\chi- M_{\t s} h^{\rm e}_\chi),
\end{gather}
\end{subequations}
where $M_{\rm s}$ is the saturation magnetization. The dynamic effective field $\mbf h^{\rm e}$ can be written as a linear combination of the dynamic magnetization, $h^{\rm e}_i = -\Lambda_{ij}m_j$ \cite{Gallardo22a}, where $\Lambda_{ij}$ is the magnetic tensor describing the self-interactions \cite{Korber21c}. 
Then, Eqs.~\eqref{LL des} result in
\begin{equation}
    \omega\begin{pmatrix}
        m_\chi\\
        m_\rho
    \end{pmatrix} = i \gamma\mu_0 M_{\t s} \tilde{\mbf T} \begin{pmatrix}
        m_\chi\\
        m_\rho
    \end{pmatrix},
\end{equation}
with the dynamic matrix
\begin{equation}
    \tilde{\mbf T} = \begin{pmatrix}
        -\Lambda_{\rho\chi} & -\Lambda_{\rho\rho} - \frac{H^{\rm e0}_{z'}}{M_{\t s}}\\
    \Lambda_{\chi\chi}+\frac{H^{\rm e0}_{z'}}{M_{\t s}} & \Lambda_{\chi\rho}
    \end{pmatrix}.
\end{equation}
By solving this eigenvalue problem, the following dispersion relation is obtained 
\begin{equation}
    f = \frac{\gamma\mu_0M_{\t s}}{2\pi}\left[i\Lambda_{\chi\rho}+\sqrt{\left(\frac{H^{\rm e0}_{z'}}{M_{\t s}}+\Lambda_{\rho\rho}\right)\left(\frac{H^{\rm e0}_{z'}}{M_{\t s}}+\Lambda_{\chi\chi}\right)}\right], 
    \label{eq: dispersion relation}
\end{equation} 
where, as shown later, it has been considered that $\Lambda_{\rho\chi}=-\Lambda_{\chi\rho}$. Eq.~(\ref{eq: dispersion relation}) maintains the structure derived in Refs. \cite{Cortes13,Otalora16,Gallardo22a,Salazar21}, where the elements $\Lambda_{\chi\rho}$ and $\Lambda_{\rho\chi}$ contain the magnetochiral properties, since they depend on the sign of the wave vector $\mathbf{k}$ and the mode index $l$. Nevertheless, it will be shown that such terms also contain the interfacial DMI so that the SW dispersion and azimuthal modes can be nonreciprocal because of exchange, DMI, and dipole-dipole interaction. According to Eq.~(\ref{eq: dispersion relation}), the $z'$ component of the static effective field ($H^{\rm e0}_{z'}$) and the linear magnetic tensor elements $\Lambda_{ij}$ are required to describe the SW dynamics of the tubular system. In the following section, these quantities are derived.  

\subsection{Magnetic effective fields and tensors}
\label{SubsecA}
%
The exchange interaction contribution to the effective field is given by $\mbf H^{\t{ex}} = l_{\t{ex}}^2\mbf\nabla^2\mbf M $, which results in $\mbf H^{\t{ex}0} = \mathcal C H_{\t{ex}} \sin\theta\hat{\mbf\varphi}$, and an exchange tensor
\begin{equation*}
\tilde{\mbf \Lambda}_{\t{ex}}=  \frac{H_{\t {ex}}}{M_{\t s}}\left(
\begin{array}{cc}
  \cos^2 \theta + l^2+  k^2 R^2\beta& i\Delta_{\t {ex}} \\
 -i\Delta_{\t {ex}} &1+l^2+k^2 R^2\beta \\
\end{array}
\right).
\end{equation*}
Here, $H_{\t{ex}} = M_{\t s}l_{\t{ex}}^2/\beta R^2$ and $\Delta_{\t {ex}} = 2l\cos\theta$, with $\beta=a/R$, being $R$ and $a$ the outer and inner radii, respectively (see Fig.~\ref{FIG1}). 
In this calculation, as in the following ones, an ultrathin nanotube is considered so that some integrals are averaged along the tube thickness.
For the effective fields, this is
\begin{equation}
    \bar{\mbf H} =  \frac{1}{R(1-\beta)}\int_{\beta R}^R \mbf H d\rho.
\end{equation}

The DM field is derived in Appendix \ref{AppA}, where its static part is $\mbf H^{\t{dm}0} =-H_{\t{dm}}\mathcal C\sin\theta
\hat{\mbf\varphi}$, and the DMI tensor is
\begin{equation}
\tilde{\mbf \Lambda}_{\t {dm}} = \frac{H_{\t{dm}}}{M_{\t s}}\begin{pmatrix} \cos^2\theta & i\Delta_{\t{dm}} \\
-i\Delta_{\t{dm}}& 1       
    \end{pmatrix},
    \label{LambdaDM}
\end{equation}
where \(H_{\t{dm}} = 2D\ln(1/\beta)/[\mu_0 M_{\t s}R(1-\beta)\)], $D$ is the Dzyaloshinskii-Moriya strength, and
\begin{equation}
    \Delta_{\t{dm}} = l\cos\theta-k\mathcal C\sin\theta \frac{R(1-\beta)}{\ln (1/\beta)}.
\end{equation} 

In flat structures, it is well known that the FM/HM interfaces exhibit a significant perpendicular anisotropy \cite{Kim17,Samardak20,Tacchi23}. Hence, in the current case, a radial uniaxial anisotropy field is considered, given by $\mbf {H}^{\rm an} = H_{\rm an}(\mbf M\cdot \hat{\rho})\hat{\rho}$, where ${H}^{\rm an}=2 K_{\rho}/\mu_0 M_{\rm s}$ with $K_{\rho}$ is the radial anisotropy constant. Therefore, 
 the associated tensor is
\begin{equation}
\tilde{\mbf\Lambda}_{\rm an} =-\frac{{H}^{\rm an}}{M_{\rm s}}\begin{pmatrix}
         0 & 0\\
         0 & 1 
    \end{pmatrix}.
\end{equation} 

For the dipolar contribution, it is noted that $\mbf \nabla \cdot \mbf M_0 = 0$ and $\mbf M_0\cdot \hat{\rho}  = 0$ on the surface of the cylindrical mantle. The surfaces at the tube ends do not contribute \cite{Landeros09,Mehlin18,Vasyukov18} because an extended NT along $z$ is assumed. Then, the static dipolar effective field is $\mbf H^{\t{dip}0} = 0$. On the other hand, the elements of the dipolar tensor $\tilde{\mbf\Lambda}_{\rm dip}$ are given by (see details in Appendix \ref{ap: dip dynamic tensor}),
\begin{align}
    (\tilde \Lambda_{\rm dip})_{\chi\chi} &=(l\cos\theta-kR_{\t m}\mathcal C\sin\theta)^2\mathcal I_0\\
    (\tilde \Lambda_{\rm dip})_{\chi\rho} &= -(\tilde \Lambda_{\rm dip})_{\rho\chi} = -i\Delta_{\rm dip} \\
    (\tilde \Lambda_{\rm dip})_{\rho\rho} &= \mathcal I_2,
\end{align}
with
\begin{equation}
     \Delta_{\rm dip} = (l\cos\theta-kR_{\t m}\mathcal C\sin\theta) \mathcal I_1.
\end{equation}
Here, an average radius $R_{\t m} = R(1+\beta)/2$ has been defined, which comes from the midpoint approach for integration along the radial direction. The functions $\mathcal I_{0,1,2}$ are integrals (in $\varphi$) that contain the modified Bessel function of the second kind \cite{Gallardo22a}. These integrals depend on the absolute value of \(l\) and \(k\) and can only be done numerically for all wave-vector ranges. 

For the magnetic configuration $\mbf M_0$ to remain stable, the resulting static effective field must align with the equilibrium magnetization. Under this condition, it is required that $\mbf H^{\rm e0}\cdot \hat{\mbf \chi} = 0$, which results in the equilibrium condition
\begin{equation}
\mathcal{C}(H_{\t{ex}}+H_{\t{dm}})\cos\theta\sin\theta +H_{\rm ze} \sin(\mathcal{C}\theta-\theta_{\rm ze})=0, \label{eq: equilibrium condition}
\end{equation}
where $\theta_{\rm ze}$ is the angle that the external field makes with respect to the $z$ axis, and $H_{\rm ze}$ is the magnitude of the external field, which is given by $\mathbf{H}^{\rm ze}=H_{\rm ze}[\sin(\mathcal{C}\theta-\theta_{\rm ze})\hat{\mbf \chi}+\cos(\mathcal{C}\theta-\theta_{\rm ze})\hat{z}']$.
Note that the external field contains the azimuthal and axial directions in order to favor the condition $ \mbf M_{\rm 0}\cdot \hat{\rho} = 0 $, i.e., the equilibrium magnetization remains in the local $z$-$\varphi$ plane. 

\subsection{Dispersion relation and frequency nonreciprocity}

With the dispersion relation shown in Eq.~\eqref{eq: dispersion relation} in terms of the effective field and magnetic tensor elements specified in the previous section, the SW frequency becomes
 \begin{widetext}
     \begin{equation}
         f= \frac{\gamma\mu_0}{2\pi}\left[(M_{\t s}\Delta_{\rm{dip}}-H_{\t{ex}}\Delta_{\rm{ex}}-H_{\t{dm}}\Delta_{\rm{dm}})+\sqrt{\left(H^{\rm e0}_{z'}+M_{\rm s}\Lambda_{\rho\rho}\right)\left(H^{\rm e0}_{z'}+M_{\rm s}\Lambda_{\chi\chi}\right)}\right],  
     \label{eq: explicit dispersion relation}
     \end{equation}
 \end{widetext}
   where
\begin{equation}
        H^{\rm e0}_{z'} = H_{\rm ze}\cos(\theta-\theta_{\rm ze})-(H_{\rm{ ex}}+H_{\rm{ dm}})\sin^2\theta,
\end{equation}
%
%
%
%
%
and the diagonal tensor elements $ \Lambda_{\rho\rho}$ and $ \Lambda_{\chi\chi}$ can be obtained from the contributing tensors in section \ref{SubsecA}. 
In this case, the only terms that 
change sign under wave vector inversion are \(\Lambda_{\chi\rho}\) and \(\Lambda_{\rho \chi}\), which satisfy the condition $\Lambda_{\rho\chi}=-\Lambda_{\chi\rho}$. 
The frequency difference under wave vector inversion is $\Delta f = f (\mathbf{k})-f(-\mathbf{k})=f (k,l)-f(-k,-l)$, and then the frequency shift is given by: 
\begin{equation}
    \Delta f  = \frac{\gamma\mu_0}{\pi}(M_{\t s}\Delta_{\rm{dip}}-H_{\t{ex}}\Delta_{\rm{ex}}-H_{\t{dm}}\Delta_{\rm{dm}}) , 
\end{equation} 
which can be separated into terms proportional to the axial ($k$) and azimuthal ($l/R$) wave vectors, \(\Delta f  =  \Delta f_k +\Delta f_l\), where
\begin{equation}
    \Delta f_k  =  \frac{\mu_0 \gamma}{\pi}kM_{\t s}\mathcal C\sin\theta\left( \frac{2D}{\mu_0 M^2_{\t s}}-R_{\t m}\mathcal I^{(k,l)}_1   \right)
    \label{Deltafk}
\end{equation}
and 
\begin{equation}
    \Delta f_l  =  \frac{\mu_0 \gamma}{\pi} \frac{l}{R}M_{\t s}\cos\theta\left(R \mathcal I^{(k,l)}_1-\frac{2D}{\mu_0 M^2_{\t s}} \frac{\ln(1/\beta)}{(1-\beta)}-\frac{2l_{\t{ex}}^2}{\beta R}   \right).
    \label{Deltafl}
\end{equation}
In Eq.~(\ref{Deltafk}), it is noted that the dipolar and DM interactions induce asymmetry in the SW bandstructure for waves propagating along the tube axis. For zero DMI, the frequency shift is generated only by the dipole-dipole interaction, which is in concordance with previous results. \cite{Otalora16,Sheka20,Gallardo22a,Salazar21,Korber22}. 
It is important to note that the function $\mathcal I^{(k,l)}_1$ depends on the absolute value of the mode index $l$, making the frequency shift $\Delta f_k$ different for $l=0$ and $l\neq 0$. 
On the other hand, the frequency shift $\Delta f_l$ [see Eq.~(\ref{Deltafl})] evidences the chiral features of the azimuthal standing waves.
One can see that dipolar, exchange, and DMI contribute to azimuthal chiral effects. Namely, the circumferential standing waves will differ for $+l$ and $-l$ as long as $\theta\neq\pi/2$. The role of exchange and dipolar interactions for $\theta\neq\pi/2$ has previously been studied \cite{Salazar21}. 
However, the role of the Dzyaloshinskii-Moriya interaction has not been considered until now. 

From Eq.~\eqref{Deltafl}, it can be concluded that the exchange contribution affects the azimuthal asymmetry in the same way as the Dzyaloshinskii–Moriya interaction. Indeed, a simple analysis allows obtaining an effective parameter
\begin{equation}
    D_{\rm ex}=   \frac{2 A_{\rm ex}}{\beta R} \frac{1-\beta}{\ln{1/\beta}}.
    \label{Dex}
\end{equation}
Thus, for the azimuthal frequency shift, the exchange and DM interactions can be combined into an effective constant $D_{\rm eff}=D+D_{\rm ex}$. In the limit $R\rightarrow \infty$, it is clear that $D_{\rm ex}=0$ because the magnetochirality induced by exchange requires a finite curvature \cite{Sheka20,Salazar21}.
On the other hand, in the same limit $R\rightarrow \infty$, $\ln(1/\beta)/(1-\beta)\rightarrow 1$, $1/R^2\rightarrow0$ and $\mathcal I^{(k,l)}_1\rightarrow 0$. Hence, using the relation $k_{\varphi}=l/R$, the frequency shift becomes $\Delta f_l= (2\gamma/\pi M_{\rm s})D k_{\varphi}$, which, at $\theta=0$, corresponds to the frequency shift obtained for DE waves propagating in a planar FM/HM structure \cite{Melcher73,Costa10,Cortes13,Moon13,Kai15,Cho15,Stashkevich15,Tacchi17b,Gallardo19BCh,Kuepferling23}. 
In the case $\theta=\pi/2$, the contribution of the DMI in the nonreciprocity is given by $\Delta f_k  =  \mathcal C\frac{2 \gamma}{\pi M_{\t s}}Dk$, which also coincides with that for DE modes in planar systems.
It is worth noting that, in the general case, the dipolar terms $\mathcal I^{(k,l)}_1$ depend on the magnitude of $k$ and are the only contribution to the frequency nonreciprocity that is nonlinear with $k$.   


\begin{figure*}
\includegraphics[width=0.8\textwidth]{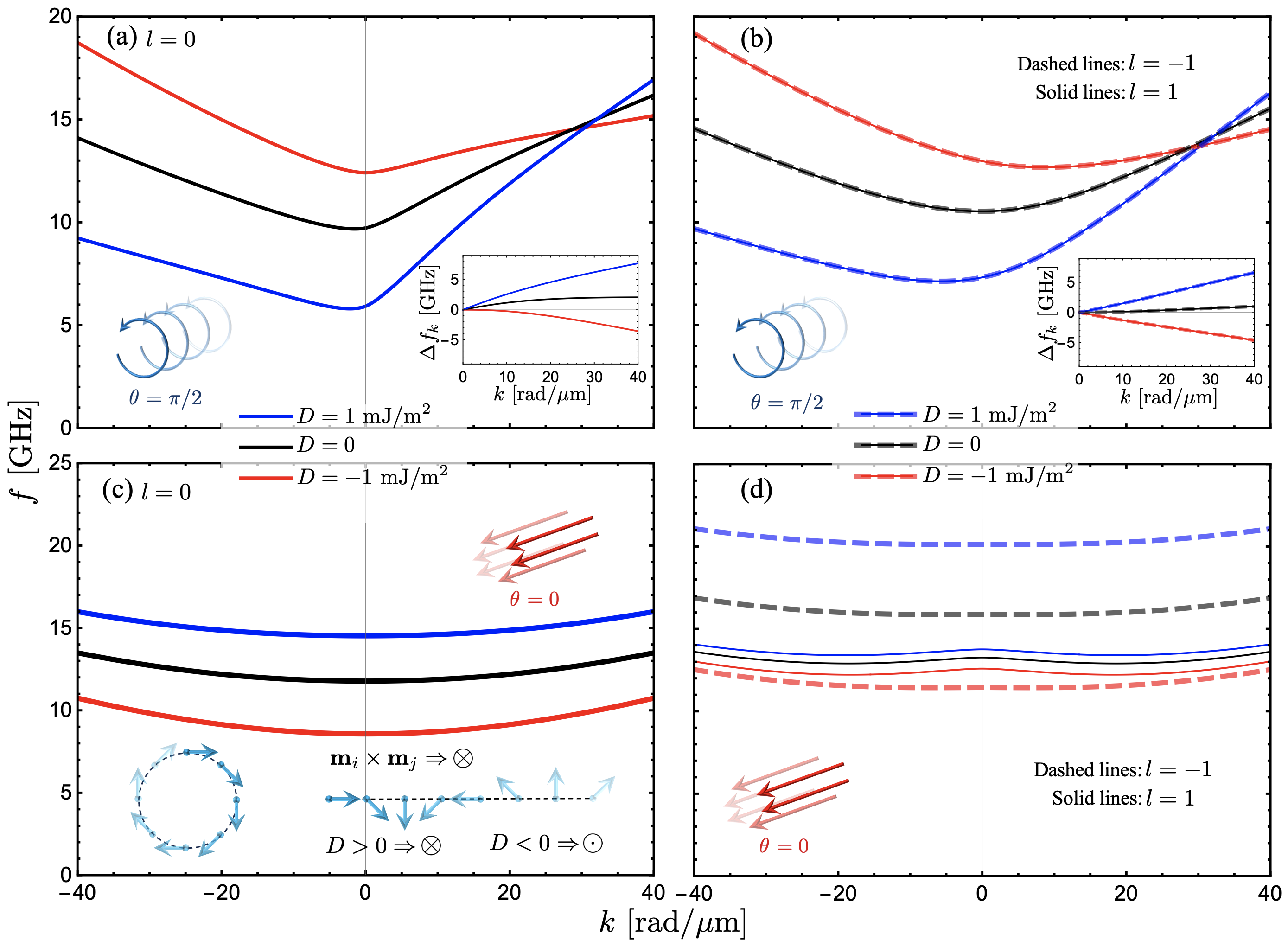}
  \caption{Spin-wave dispersions for $D=-1\ \rm mJ/m^2$ (red), $D=0$ (black) and $D=1\ \rm mJ/m^2$ (blue). In (a) and (b), the equilibrium magnetization presents a vortex-like ground state ($\theta=\pi/2$), where the case $l=0$ [$l=\pm1$] is illustrated in (a) [(b)]. In (c) and (d), the nanotube is saturated with an external field applied along its axis ($\theta=0$), where the cases (c) $l=0$ and (d) $l=\pm1$ are depicted. The bottom scheme in (c) describes the dynamic magnetization for the dynamic state $l=0$. The directions of vectors $\mathbf{m}_i\times \mathbf{m}_j$, with $\mathbf{m}_{i,j}$ being the dynamic magnetization components, and the DM vector (for $D>0$ and $D<0$) are illustrated.}
  \label{fig: Spin-waves}
\end{figure*}

\begin{figure}[t]
  \includegraphics[width=.44\textwidth]{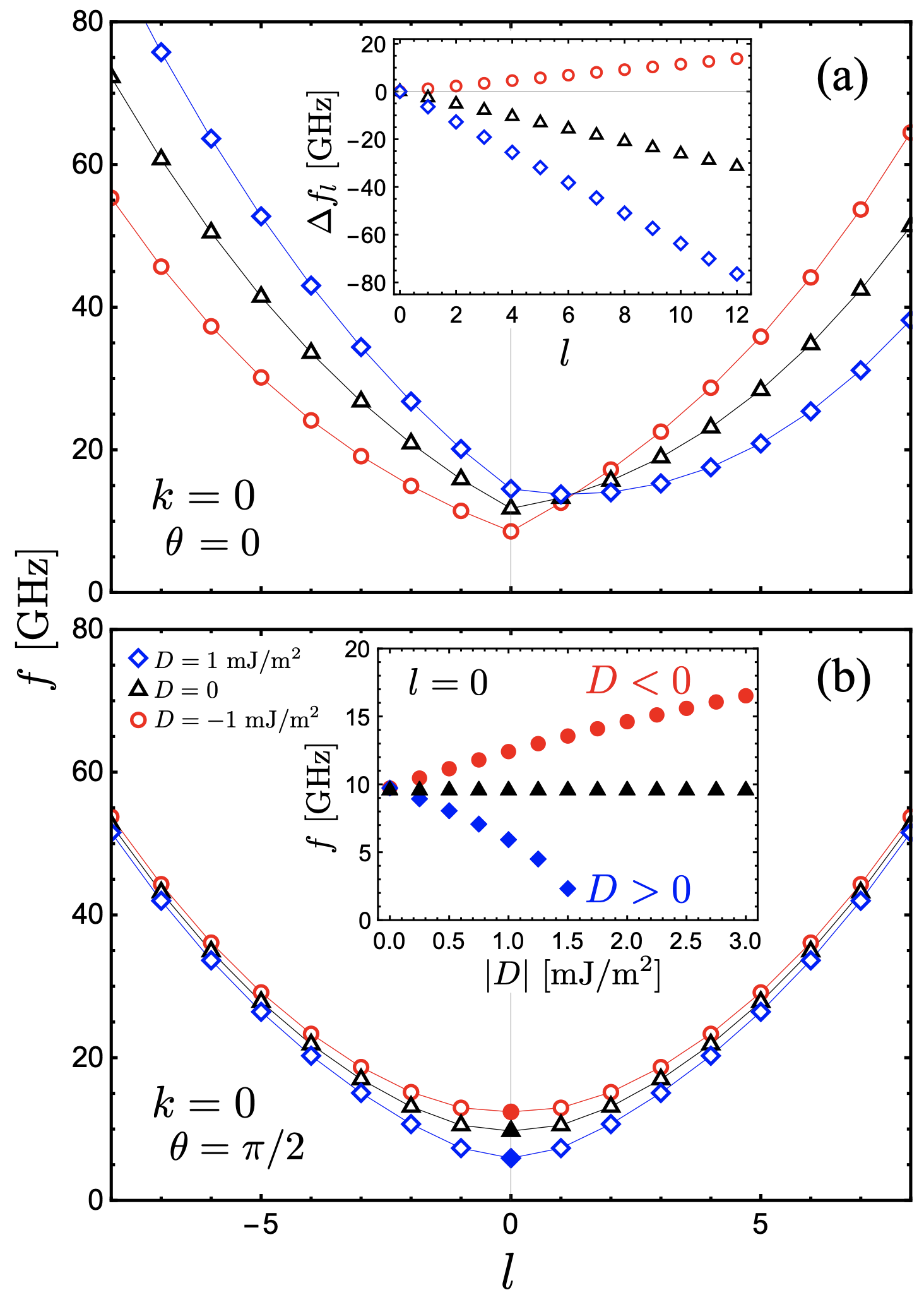}
  \caption{Spin-wave frequency as a function of the azimuthal mode index $l$ for $k=0$. Diamonds, triangles, and circles depict the cases $D=1 $ mJ/m$^2$,  $D=0$, and $D=-1 $ mJ/m$^2$, respectively. In (a), a saturated equilibrium state along $z$ is considered ($\theta=0$), where the inset shows the frequency shift $\Delta f_l$ for the cases $D=0$ and $D=1\pm $ mJ/m$^2$. The circumferential equilibrium state ($\theta=\pi/2$) is assumed in (b), where the frequency against the magnitude of the Dzyaloshinskii-Moriya constant $D$ is shown in its inset for $l=0$ (solid markers).
  } 
  \label{FIG3}
\end{figure}

\begin{figure*}[ht]
\includegraphics[width=0.95\textwidth]{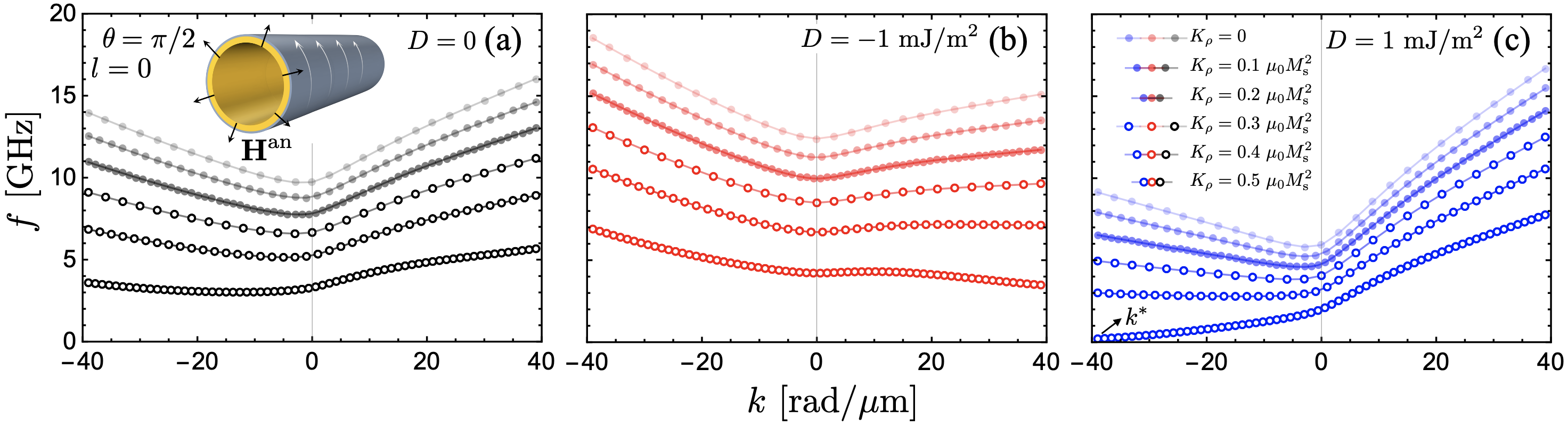}
  \caption{Spin-wave dispersions calculated for (a) $D=0$, (b) $D=-1\ \rm mJ/m^2$, and (c) $D=1\ \rm mJ/m^2$ (blue). In all cases, the vortex-like equilibrium state ($\theta=\pi/2$) and $l=0$ are considered. The plots are calculated for different values of the radial anisotropy constant $K_{\rho}$, as depicted in (c). The filled circles consider the cases $K_{\rho}=0$, $K_{\rho}=0.1\ \mu_0 M_{\rm s}^2$, and  $K_{\rho}=0.2\ \mu_0 M_{\rm s}^2$. While the open circles are calculated for $K_{\rho}=0.3\ \mu_0 M_{\rm s}^2$, $K_{\rho}=0.4\ \mu_0 M_{\rm s}^2$, and  $K_{\rho}=0.5\ \mu_0 M_{\rm s}^2$. In (c), the wave vector $k^*$, at which the frequency becomes zero, is highlighted.}
  \label{FIG4}
\end{figure*}

\subsection{Small wave-vector approximation}

Analytical expressions for the integrals $\mathcal I^{(k,l)}_i$ can be derived at the small wave-vector limit. Such expressions are obtained up to zeroth order in Appendix \ref{ap: dip dynamic tensor}. By focusing on the frequency asymmetry, the frequency shift at $k\rightarrow 0$ becomes
\begin{equation}
    \Delta f_k^{k\rightarrow 0}  \approx  \frac{\mu_0 \gamma}{\pi}k M_{\t s}\mathcal C\sin\theta\left(\frac{2D}{\mu_0 M^2_{\t s}}+\frac{R-a}{2}\delta_{l,0}    \right).
    \label{DeltafkAn}
\end{equation}
Note that for $l=0$, the DM and dipolar couplings induce frequency nonreciprocity.
Nonetheless, for $l\neq 0$, the frequency asymmetry is due to only DMI because, in this case, $\mathcal I^{(k,l)}_1(k\rightarrow 0)=0$. Also, for $D=0$, Eq.~(\ref{DeltafkAn}) coincides with Eq.~(18) of Ref. \cite{Gallardo22a} when the equilibrium magnetization is in a circumferential vortex state ($\theta=\pi/2$).

On the other hand, the small-\(k\) approximation for the frequency asymmetry associated with the azimuthal standing waves is 
\begin{equation}
    \Delta f_l^{k\rightarrow 0}  \approx  -\frac{\mu_0 \gamma}{\pi}\frac{l}{R}M_{\t s}\cos\theta\left(\frac{2D\ln(1/\beta)}{\mu_0 M^2_{\t s}(1-\beta)}+\frac{2l_{\t{ex}}^2}{\beta R}  \right).
    \label{DeltaflAn}
\end{equation}
Hence, in the limit $k \rightarrow 0$, only the exchange interaction and DMI induce magnetochirality if $l\neq0$. In other words, Eq.~(\ref{DeltaflAn}) suggests that the asymmetry of azimuthal standing waves is predominantly due to the exchange interaction and DMI, and the dipolar interaction plays a minor role. Nonetheless, it is worth noticing that only the dipolar interaction can induce nonlinear effects in the frequency shift, as evidenced in the inset of Fig.~\ref{fig: Spin-waves}(a). This effect is explained by the dependence of the integral \(\mathcal I^{(k,l)}_1\) on \(k\), which is lost in the zeroth order approximation considered in Eqs.~\eqref{DeltafkAn} and \eqref{DeltaflAn} and higher order terms are needed. 
Note that these analytical expressions [Eqs.~(\ref{DeltafkAn}) and (\ref{DeltaflAn})] account for asymmetry in both $k$ and $l$, providing insights into the role of curvature and the various interactions in SW dynamics.

\section{Results}
In the subsequent discussion, magnetic parameters for Permalloy (Ni$_{80}$Fe$_{20}$) are used for the calculations, namely $M_{\rm{s}} = 796$~kA/m and $l_{\rm{ex}} = 5.7$~nm (equivalent to an exchange constant $A_{\rm{ex}}=13$~pJ/m). The gyromagnetic ratio is $\gamma = 175.929$~GHz/T. In addition, the geometrical parameters of the tubular system are $R=40$ nm and $a= 35$ nm, so the NT thickness is 5 nm. 

Fig.~\ref{fig: Spin-waves} shows the spin-wave dispersion for the cases (a-b) $\theta = \pi/2$ and (c-d) $\theta = 0$, evaluated for $K_{\rho} = 0$. To stabilize saturated and vortex-like equilibrium states, it is assumed that an external field of magnitude $\mu_0 H_{\rm ze} = 130$~mT is applied along the local $z'$ direction, ensuring $\theta = \theta_{\rm ze}$, with $\theta_{\rm ze} = 0$ for the saturated state and $\theta_{\rm ze} = \pi/2$ for the circumferential equilibrium configuration.
Fig.~\ref{fig: Spin-waves}(a) displays the dispersion at $\theta = \pi/2$, for $D = -1~\mathrm{mJ/m^2}$ (red), $D = 0$ (black), and $D = 1~\mathrm{mJ/m^2}$ (blue). The asymmetry in the SW dispersion at $D = 0$ arises from the dynamic dipolar interaction, which causes counterpropagating waves to have different wavelengths at the same frequency \cite{Otalora16,Sheka20,Gallardo22a,Salazar21,Korber22}. As indicated by Eq.~(\ref{Deltafk}), the exchange interaction does not contribute to this frequency asymmetry.
When the DMI is active, the dispersive SW properties are modified depending on the sign of $D$ and the helicity of the curling magnetization. For a counterclockwise ($\mathcal{C}=1$) vortex magnetization and for $D > 0$, the DMI reinforces the asymmetry induced by the dipolar interaction, whereas for $D < 0$, the DMI opposes the dipolar-induced asymmetry. This results in a negative frequency shift $\Delta f_k$, as illustrated in the inset of Fig.~\ref{fig: Spin-waves}(a).
For the states $l = \pm 1$ (at $\theta = \pi/2$), the azimuthal standing waves are degenerate in frequency because $\Delta f_l=0$. This result holds for any $l\neq0$.  
Specifically, the SW dispersion depends on the magnitude of $l$ [see Fig.~\ref{fig: Spin-waves}(b)]. In this case, the frequency asymmetry (see inset) differs from the case where $l = 0$. This behavior is attributed to the dipolar interaction, where the dependence of $\Delta f_k$ on $l$ is determined by the integral $\mathcal{I}_1$ in Eq.~(\ref{Deltafk}). The analytical expression in Eq.~(\ref{DeltafkAn}), evaluated at small wave vectors, explicitly reveals this dependence. Indeed, at $D=0$, the frequency shift tends to zero in the limit $k\rightarrow 0$, as shown in the inset of Fig.~\ref{fig: Spin-waves}(b).

For the saturated case ($\theta = 0$), the SW dispersion is symmetric because $\Delta f_k \propto \sin\theta$. Neither dipolar interactions nor DMI induce nonreciprocity with respect to $k$, as shown in Figs.~\ref{fig: Spin-waves}(c) and \ref{fig: Spin-waves}(d). Nevertheless, in this configuration, the DMI shifts the magnon frequency depending on the sign of $D$. As illustrated in the scheme at the bottom of Fig.~\ref{fig: Spin-waves}(c), the sign of the Dzyaloshinskii-Moriya constant $D$ determines whether the DM energy $\epsilon_{\mathrm{dm}}$ is maximized or minimized. For $D > 0$, $\epsilon_{\mathrm{dm}} \propto \mathbf{D} \cdot (\mathbf{m}_i \times \mathbf{m}_j)$ reaches a maximum because $\mathbf{D}$ and the dynamic vector spin chirality $\mathbf{m}_i \times \mathbf{m}_j$ are parallel \cite{Grohol05,Onoda07,Menzel12,Ishizuka18}. 
Conversely, for $D < 0$, $\mathbf{D}$ points out of the page, resulting in $\mathbf{D} \cdot (\mathbf{m}_i \times \mathbf{m}_j) < 0$, which reduces $\epsilon{_\mathrm{dm}}$. Consequently, the frequency decreases (increases) for $D < 0$ ($D > 0$). 
In Fig.~\ref{fig: Spin-waves}(d), it is evident that the modes depend on the sign of $l$, unlike the vortex-like case. This dependence is associated with the finite frequency shift $\Delta f_l$ that is maximized for $\theta=0$, where dipolar, exchange, and DMI interactions contribute to a dependence proportional to $l \cos\theta$, as derived in Eq.~(\ref{Deltafl}).

The asymmetry associated with the azimuthal index $l$ is further analyzed in Fig.~\ref{FIG3}, where the frequency as a function of $l$ is plotted for zero axial wave vector. In Fig.~\ref{FIG3}(a), the three cases $D=-1\ \rm mJ/m^2$ (circles), $D=0$ (triangles), and $D=1\ \rm mJ/m^2$ (diamonds) are shown for an equilibrium magnetization aligned along the tube axis $\theta=0$. It is observed that in this configuration, the curve at $D=0$ exhibits a degree of nonreciprocity concerning the change in sign of $l$, wherein the frequency minimum occurs at positive values [$l=1$ in Fig.~\ref{FIG3}(a)]. This behavior is consistent with previously reported results \cite{Salazar21}, where this asymmetry is induced by dipolar and exchange interactions, as described in Eq.~(\ref{Deltafl}).
When the interfacial DMI is included, the frequency asymmetry is enhanced for $D>0$, shifting the minimum to higher values of $l$. Conversely, for $D<0$, the frequency minimum shifts to negative $l$ values. As illustrated in the inset of Fig.~\ref{FIG3}(a), the deviation of $\Delta f_l$ around the $D=0$ curve (triangles) is symmetric for positive and negative values of $D$, namely $\vert \Delta f_l (D>0)-\Delta f_l (D=0)  \vert=\vert \Delta f_l (D<0)-\Delta f_l (D=0)  \vert$. This behavior is associated with the term in $\Delta f_l$ that is linear with $D$.

In the vortex configuration ($\theta=\pi/2$), the frequency depends on the magnitude of $l$, resulting in a symmetric behavior against $l$, as shown in Fig.~\ref{FIG3}(b). The parabolic dependence highlights the dominance of the exchange interaction, which scales as $\sim l^2$. Similar to the saturated case, the DMI either increases or decreases the frequencies depending on the sign of $D$. This is further demonstrated in the inset of Fig.~\ref{FIG3}(b), where the frequency at $l=0$ is plotted as a function of $\vert D \vert$. As the magnitude of $D$ increases, the modes for $D<0$ increase their frequency, while those for $D>0$ are excited at a lower frequency. Indeed, for $D\approx 1.6\ \rm mJ/m^2$, the frequency is close to zero, leading to the emergence of soft (or Goldstone) modes indicating the destabilization of the equilibrium state \cite{Leaf06,Montoncello07,Rios22,Grassi22,Kisielewski23}. If a given mode with $l\neq 0$ reaches zero frequency before the mode at $l=0$, the new ground state is expected to exhibit a magnetic texture with an azimuthal modulation determined by $l$. The analysis of different ground states in tubular systems with Dzyaloshinskii-Moriya interaction is beyond the scope of this work and will be addressed in a forthcoming publication \cite{Mimica25}.

\begin{figure}[t]
\includegraphics[width=0.48\textwidth]{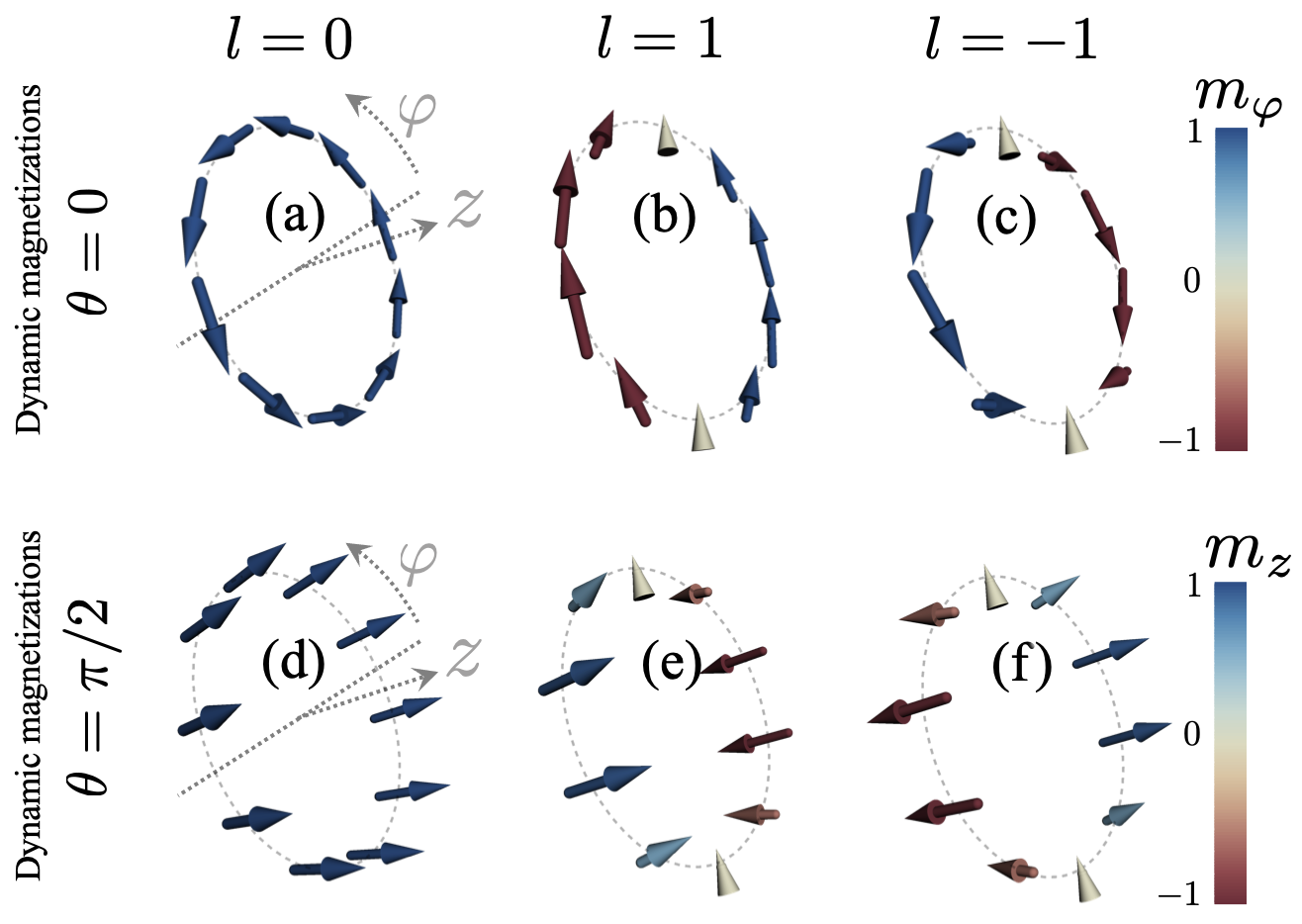}
  \caption{Dynamic magnetization calculated for $D=0$ and $K_{\rho}=0$. In (a)-(c), the dynamic magnetizations as a function of $\varphi$ are shown for the dynamic states $l=0,\pm1$ with $\theta=0$. In (d)-(f), the vortex-like configuration is illustrated ($\theta=\pi/2$) where the dynamic magnetization are evaluated for $l=0,\pm1$.}
  \label{FIG5}
\end{figure}

The impact of radial anisotropy is illustrated in Fig.~\ref{FIG4}, where the spin-wave dispersion is analyzed for $l=0$, $\theta=\pi/2$, and varying values of the anisotropy constant $K_{\rho}$. Dispersion curves are presented for three cases: $D=0$, $D=-1\ \rm mJ/m^2$, and $D=1\ \rm mJ/m^2$, corresponding to Figs.~\ref{FIG4}(a), \ref{FIG4}(b), and \ref{FIG4}(c), respectively.
For $D=0$, the dispersion curves reveal a progressive reduction in frequency as $K_{\rho}$ increases, while the frequency shift ($\Delta f_k$) remains unaffected by changes in anisotropy. A notable feature is the shift of the dispersion minimum to a finite wave vector $k^*$, indicating the existence of a critical anisotropy constant $K^{*}_{\rho}$ where the frequency becomes zero, i.e., $f(k^*)=0$. Beyond this critical anisotropy value, the nanotube is predicted to develop a textured ground state with periodicity along the $z$ axis, characterized by a wavelength $\lambda=2\pi/k^*$. This idea aligns with results obtained in other systems \cite{Leaf06,Montoncello07,Rios22,Grassi22,Kisielewski23}, where the critical wave vector $k^*$ enables the prediction of a periodic magnetic texture.
In Fig.~\ref{FIG4}(a), for $D=0$, the minimum occurs at $k^*<0$, reflecting the chiral properties inherent to the emerging magnetic texture. When the DMI is introduced, the $f(k^*)=0$ state is observed at $k^*>0$ for $D=-1\ \rm mJ/m^2$, and at $k^*<0$ for $D=1\ \rm mJ/m^2$, as shown in Figs.~\ref{FIG4}(b) and \ref{FIG4}(c), respectively. For the specific case of $D=1\ \rm mJ/m^2$ and $K_{\rho}=0.5\ \mu_0 M_{\rm s}^2$, which means that radial anisotropy and dipolar fields are balanced, the zero-frequency state is nearly reached at $k^*=-40$~rad/$\mu$m, suggesting the formation of a magnetic texture with a similar pitch vector, as occurs in planar films \cite{Rios22,Grassi22,Kisielewski23}.

The periodic equilibrium states of thin films with DMI have been previously reported \cite{Rios22}. In this context, the predicted textures in nanotubes based on dynamic aspects align with results reported in Ref.~\onlinecite{Yershov20}, where periodic equilibrium states are observed at a critical value of the DM constant.
This dynamic prediction of periodic magnetic textures under critical conditions is nontrivial. Intuitively, one might expect that increasing anisotropy would stabilize a purely radial magnetization orientation in the nanotube. However, the results demonstrate that even for $D=0$, the tube exhibits a texture with a periodic radial component, where the chirality arises from dipolar interactions.
For dynamic states evaluated at $l\neq 0$, a similar trend to that discussed in Fig.~\ref{FIG4} is observed (not shown). In the saturated state, anisotropy also reduces the frequency of the modes. However, in this case, the frequency minima occur at zero wave vector, indicating that anisotropy uniformly enhances the radial component of the equilibrium magnetization across the entire nanotube.

Finally, the spin-wave profiles are calculated for $D=0$ and $K_{\rho}=0$. Fig.~\ref{FIG5} illustrates the normalized dynamic magnetization (in arbitrary units) for the dynamic states $l=0,\pm1$. The dynamic magnetization component in the tube mantle is highlighted using a color code.
For $l=0$, the dynamic magnetization is shown in Fig.~\ref{FIG5}(a), exhibiting the same amplitude along the $\varphi$-direction. In contrast, the case $l=1$ [see Fig.~\ref{FIG5}(b)] presents an almost uniform dynamic state, where the dynamic magnetization tends to align in a given direction. This configuration clearly implies a reduction in the exchange energy, as it promotes a parallel orientation of the magnetic moments.
By comparing the state $l=1$ with the one illustrated in Fig.~\ref{FIG5}(c) ($l=-1$), it is seen that the dynamic energies are different, with the configuration $l=-1$ being more energetic than $l=1$. This observation explains the asymmetry of the exchange energy evident in Fig.~\ref{FIG3}(a) (see black triangles).

When the DMI is included, for the case $\theta=0$, the energy in the configuration $l=-1$ is either reduced or increased depending on the sign of $D$. 
For $D>0$, the direction of the DM vector $\mathbf{D}$ between two neighboring spins across the nanotube's perimeter is determined by $+\hat{z}$, while the product $\mathbf{m}_i\times \mathbf{m}_j$ in the configuration shown in Fig.~\ref{FIG5}(c) also points along $+z$.
Consequently, the dynamic energy (or frequency) increases for the case $l=-1$, as the DM energy [$\epsilon_{\rm dm}\propto \mathbf{D}\cdot(\mathbf{m}_i\times \mathbf{m}_j)$] increases.
In contrast, for $D<0$, the dynamic DM energy becomes negative, and thus, the state with $l=-1$ becomes less energetic than the state with $l=1$, consistent with the results shown in Fig.~\ref{FIG3}(a) (see red open circles). It is worth mentioning that the dynamic magnetization vectors illustrated in Figs.~\ref{FIG5}(a)--(c) are slightly modified in the presence of DMI, at least within the range $\vert D\vert <1$~mJ/m${^2}$. Therefore, although the states illustrated in Fig.~\ref{FIG5} are not equivalent under the presence of DMI, they allow understanding why the frequency of the modes increases or decreases when DMI is active.
In the vortex configuration ($\theta=\pi/2$), the state $l=0$ [see Fig.~\ref{FIG5}(d)] exhibits a dynamic magnetization distribution with the same amplitude along the $\varphi$-direction, as in the case $\theta=0$. In this configuration, the magnetization component $m_z$ lies along the tube mantle. The cases $l=\pm 1$ are shown in Figs.~\ref{FIG5}(e)--(f), where the dynamic energy is identical for these states. This result implies that the frequencies of states with negative and positive $l$ are equal, thereby supporting the results presented in Fig.~\ref{FIG3}(b).

\section{Summary}

The theoretical study provides a detailed analysis of spin-wave dynamics in nanotubes, emphasizing the crucial roles of Dzyaloshinskii-Moriya interaction and curvature. It is shown that nonreciprocity in spin-wave propagation is a key characteristic of these systems, arising from a combination of exchange, DMI, and dipolar interactions. In the vortex-like configuration, a frequency nonreciprocity is observed wherein, for a positive DM constant, the DMI reinforces the asymmetry induced by the dipolar interaction, while in the contrary case, the asymmetry caused by the DMI is opposed to the dipolar one. In a longitudinal magnetization state, the exchange and DM interactions predominantly lead to chiral effects in the SW bandstructure for azimuthal standing waves. In contrast, propagation along the tube axis is reciprocal in frequency. Under the inclusion of a radial anisotropy, it is found that it reduces the frequency of the modes and shifts the dispersion minimum to a finite wave vector for the vortex state. For the saturated state, the frequency minima occurs at zero wave vector. The properties of the spin-wave modes are close to zero frequency, which allows for the prediction of the emergence of chiral magnetic textures. The insights from this work can be applied to the development of three-dimensional chiral magnetic nanostructures, offering new possibilities for magnonics and nanoscale information processing.

\section*{ACKNOWLEDGMENTS}
The authors acknowledge financial support from Fondecyt Grants 1250803 and 1201153, and Basal Program for Centers of Excellence, Grant AFB220001 CEDENNA, ANID. Fruitful discussions with Felipe Brevis are highly recognized. 

\appendix 
\section{Effective Dzyaloshinskii-Moriya field}
\label{AppA}
The effective Dzyaloshinskii-Moriya field must be carefully calculated to incorporate curvature effects. In cylindrical geometry, the interfacial DM energy term is expressed as \cite{Mancilla20},
\begin{equation}
    E_{\t{dm}}(\mbf M ) = \frac{D}{M_{\t{s}}^2}\int_{\Omega}\hat{\mbf\rho}\cdot[(\mbf M (\mbf \nabla\cdot\mbf M)-(\mbf M\cdot\mbf\nabla)\mbf M]dV.
\end{equation}
Due to the non-commutativity between the unit vector $\hat{\mbf \rho}$ and the operator $\mbf \nabla$, curvature-induced terms naturally emerge in the effective DMI field.
This field can be obtained by calculating the functional derivative of the energy with respect to the magnetization. This is 
\begin{align}
    \delta E_{\rm dm}(\mbf M,\mbf v) &\equiv \lim_{\epsilon\to 0} \frac{E_{\rm dm}(\mbf M+\epsilon \mbf v)-E_{\rm dm}(\mbf M)}{\epsilon} \notag\\
    &= \int_{\Omega} \frac{\delta \epsilon_{\rm dm}}{\delta \mbf M} \cdot\mbf v dV
    \label{deltaE1}
\end{align}
for an arbitrary vector $\mbf v$. Then, the effective DM field is
\begin{equation}
    \mbf{H}^{\t{dm}}= -\frac{1}{\mu_0} \frac{\delta \epsilon_{\t{dm}}}{\delta \mbf M},
\end{equation}
with \(\epsilon_{\rm{dm}}\) being the DM energy density ($\epsilon_{\rm dm}=\partial E_{\rm {dm}}/\partial V $). After some algebraic manipulations, the following expression for the effective field is obtained:
\begin{gather}
    \mbf{H}^{\t{dm}} = -\frac{2D}{\mu_0 M^2_{\t s}}\left(\hat{\mbf\rho}(\mbf\nabla\cdot\mbf M)-\mbf\nabla(\hat{\mbf\rho}\cdot\mbf M)+\frac{M_\varphi\hat{\mbf\varphi}}{\rho}\right).
    \label{heff dmi}
\end{gather}
Note that a thin nanotube is considered in this work. Therefore, the expression for a surface contribution in Eq.~(\ref{deltaE1}) has been omitted. Finally, by considering the dynamic and static magnetization components, it is easy to obtain $\mbf H^{\t{dm}0} =-H_{\t{dm}}\sin\theta
\hat{\mbf\varphi}$ and $\mbf h^{\t{dm}}=-\tilde{\mbf \Lambda}_{\t {dm}} \mbf m$, where matrix $\tilde{\mbf \Lambda}_{\t {dm}}$ is given in Eq.~(\ref{LambdaDM}).
\\

\section{Effective dipolar field}
\label{ap: dip dynamic tensor}

The dipolar field is derived in a manner similar to that described by Alvarado et al. \cite{Alvarado19}. The dipolar effective field can be obtained as $\mbf H^{\t {dip}} = -\mbf\nabla\Phi(\mbf r,t)$, where the magnetostatic potential $\Phi$ is given by
\begin{equation}
    \Phi(\mbf r,t) = -\int_{\Omega} \frac{\mbf\nabla' \cdot \mbf M(\mbf r',t)}{4 \pi|\mbf r-\mbf r'|}dV' + \int_{\partial \Omega} \frac{\hat{\mbf n}'\cdot\mbf M(\mbf r',t)}{4\pi |\mbf r-\mbf r'|}dS'. \label{mag potential}
\end{equation}
By considering the magnetization described in section \ref{Theoretical model}, the volumetric and surface magnetic charges are 
\begin{equation}
        \mbf \nabla\cdot\mbf M = \left[\frac{m_\rho}{\rho} + i\left(k \sin\theta-\frac{l\cos\theta }{\rho}\right)m_{\chi}\right]e^{i(kz+l\phi-\omega t)}
\end{equation}
and 
\begin{equation}
        \mbf n\cdot\mbf M = \pm m_\rho e^{i(kz+l\phi-\omega t)},
\end{equation}
respectively. Therefore, the tensor elements of $\tilde{\mathbf{\Lambda}}_{\rm dip}$ are 
\begin{widetext}
\begin{subequations}\label{lambda dip}
\begin{align}
    (\Lambda_{\rm dip})_{\chi\chi} &=\iiint(l\cos\theta-k\rho\sin\theta)(l\cos\theta-k\rho'\sin\theta)\frac{1}{\rho}G_0(\rho,\rho',\varphi')d\rho d\rho'd\varphi'\\
     (\Lambda_{\rm dip})_{\chi\rho} &=  -i\iiint \left(l\cos\theta-k\rho\sin\theta\right)\frac{\rho'}{\rho}G_1(\rho',\rho,\varphi')d\rho d\rho'd\varphi' \\
    (\Lambda_{\rm dip})_{\rho\chi} &=  i\iiint  (l\cos\theta-k\rho'\sin\theta)G_1(\rho,\rho',\varphi')d\rho d\rho'd\varphi'\\
    (\Lambda_{\rm dip})_{\rho\rho} &= \iiint \rho'G_2(\rho,\rho',\varphi') d\rho d\rho'd\varphi'
\end{align}
\end{subequations}
where, 
\begin{subequations}
\begin{align}
G_0(\rho,\rho',\varphi') &= \frac{e^{il\varphi'}}{2\pi R(1-\beta )} K_0\left(\vert k \vert \sqrt{\rho^2+\rho'^2-2\rho\rho'\cos\varphi'}\right)\\
G_1(\rho,\rho',\varphi') &=  \partial_\rho G_0(\rho,\rho',\varphi ')\\
G_2(\rho,\rho',\varphi') &=  \partial_{\rho\rho'} G_0(\rho,\rho',\varphi ').
\label{G function}
\end{align}
\end{subequations}
\end{widetext}
Here, $K_0$ is the zeroth modified Bessel function of the second kind. The integrals given in Eq.~(\ref{lambda dip}) must be solved numerically. Notice that, unlike exchange and DM dynamic tensors, the dipolar interaction has the condition \((\Lambda_{\rm dip})_{\rho\chi}\neq -(\Lambda_{\rm dip})_{\chi\rho}\). Nevertheless, because a very thin nanotube is considered, the integrals in \(\rho\) and \(\rho'\) can be approximated as \( \int_{0}^d F(x) dx \approx d F(d/2) \). Thus, 
\begin{subequations}
\begin{align}
    (\tilde \Lambda_{\rm dip})_{\chi\chi} &=(l\cos\theta-kR_{\t m}\sin\theta)^2\mathcal I^{(k,l)}_0\\
    (\tilde \Lambda_{\rm dip})_{\chi\rho} &=   -i(l\cos\theta-kR_{\t m}\sin\theta) \mathcal I^{(k,l)}_1\\
    (\tilde \Lambda_{\rm dip})_{\rho\rho} &= \mathcal I^{(k,l)}_2,
\end{align}
\label{SUB9}
\end{subequations}
\noindent
and $ (\tilde \Lambda_{\rm dip})_{\rho\chi}=-(\tilde \Lambda_{\rm dip})_{\chi\rho}$.
Here, \(R_{\t m} = R(1+\beta)/2\) is the midpoint between the internal and external radius of the tube. The integrals $\mathcal I^{(k,l)}_{0,1,2}$ become 
\begin{subequations}
    \begin{align}
        \mathcal I^{(k,l)}_0 &= R^2(1-\beta)^2\int \frac{1}{R_{\t m}} G_0(R_{\t m},R_{\t m},\varphi')d\varphi'\\
        \mathcal I^{(k,l)}_1 &=R^2(1-\beta)^2\int G_1(R_{\t m},R_{\t m},\varphi')d\varphi'\\
        \mathcal I^{(k,l)}_2 &=R^2(1-\beta)^2\int R_{\t m} G_2(R_{\t m},R_{\t m},\varphi')d\varphi'.
    \end{align}
\end{subequations}
These expressions require numerical integration on the azimuthal angle $\varphi'$. Nevertheless, these integrals can be analytically calculated in the case of small wave vectors. The series expansion of the Bessel function is used \(K_0(x) = -\gamma-\ln(x/2)\), where \(\gamma\) is the Euler-Mascheroni constant. Such a small $k$ approach allows for an analytical solution of the \(\varphi'\) coordinate integration using the residue theorem. An integral of the form,
\begin{equation} \label{int 02pi}
    I = \int_0^{2\pi} f(\sin\varphi',\cos\varphi')d\varphi'
\end{equation}
can be solved by using the change of variable \(z = e^{i\varphi'}\). The result is computed by calculating the residues inside the unit circle,
\begin{align}
    I &= -i \oint_{\partial  C}f\left(\frac{z-z^{-1}}{2i},\frac{z+z^{-1}}{2}\right)\frac{dz}{z}\\
    &= 2\pi \sum_C \text{res}(f/z) \label{res theo ints 0 2pi}
\end{align}
where \(\partial C\) is the circumference of radius \(r = 1\). The expressions obtained for the special case \(l = 0\),
\begin{subequations}
    \begin{align}
        \mathcal I^{(k,0)}_0 =&\  \frac{4(1-\beta)}{(1+\beta)^2}
            \left(\gamma+\ln\left(\frac{|k|R}{2}\right)\right)\\
        \mathcal I^{(k,0)}_1 =&\ -\left(\frac{1-\beta}{1+\beta}\right) 
        \\
        \mathcal I^{(k,0)}_2 =&\ 1,
    \end{align}
\end{subequations}
while for $l \neq 0$,
\begin{subequations}
    \begin{align}
        \mathcal I^{(k,l)}_0 = &\  
 \frac{1}{|l|}\left(\frac{1-\beta}{1+\beta}\right)\\
        \mathcal I^{(k,l)}_1 = &\ 0\\
        \mathcal I^{(k,l)}_2 = &\ 
            \frac{1}{1+\beta}\left[\left(\frac{1+\beta}{2}\right)^{|l|}+\beta\left(\frac{2\beta}{1+\beta}\right)^{|l|}\right].
    \end{align}
\end{subequations}

\sloppy




%

\end{document}